\documentclass[%
 reprint,
 amsmath,amssymb,
 aps,
]{revtex4-1}

\usepackage{graphicx}% Include figure files
\usepackage{dcolumn}% Align table columns on decimal point
\usepackage{bm}% bold math

\begin{document}

\preprint{APS/123-QED}

\title{Cosmological dynamics of a hybrid chameleon scenario}% Force line breaks with \\

\author{Kourosh Nozari}
\homepage{knozari@umz.ac.ir}
\author{Narges Rashidi}
\homepage{n.rashidi@umz.ac.ir}%

\affiliation{Department of Physics, Faculty of Basic Sciences, University of Mazandaran,\\
P. O. Box 47416-95447, Babolsar, IRAN}

\begin{abstract}
We consider a hybrid scalar field which is non-minimally coupled to
the matter and models a chameleon cosmology. By introducing an
effective potential, we study the dependence of the effective
potential's minimum and hybrid chameleon field's masses to the local
matter density. In a dynamical system technique, we analyze the
phase space of this two-field chameleon model, find its fixed points
and study their stability. We show that the hybrid chameleon
domination solution is a stable attractor and the universe in this
setup experiences a phantom divide crossing.
\begin{description}
\item[PACS numbers]
98.80.Cq,\, 95.36.+x
\item[Key Words]
Chameleon Cosmology; Hybrid Scalar Field; Dynamical System; Cosmic
Acceleration
\end{description}
\end{abstract}
\maketitle

\section{Introduction}
Recent cosmological observations have revealed that our universe is
currently undergoing an accelerating phase of expansion
\cite{Rie98,Per99,Rie04,Ast06,Woo07,Spe07,Hin07,Col01,Teg04,Col05,Spr06,Bou04,McE07,Kom09}.
This transition to the accelerating phase has been occurred in the
recent cosmological past. Several approaches have been proposed to
explain this late-time acceleration of the universe. Introducing
some sort of unknown energy component (dubbed ``\textit{dark
energy}''), with negative pressure, is one of these various
approaches
\cite{Cop06,Noj04,Cal02,Ark02,Pia04,Wei06,Vik05,Ani05,Wan05,Noj05,Noj06,Eli05,Zha06a,Apo06,Ala04,Nes05,lib07,Sch08,Bri07,Sam09,Cal09,Sah04,Sah06}.
The simplest candidate for dark energy is the cosmological constant.
However, the cosmological constant suffers from serious problems
such as huge amount of fine-tuning required for its magnitude and
other theoretical problems such as unknown origin and lake of
dynamics \cite{Pad03,Car01,Sah00,Pee03}. Nevertheless, the standard
cosmological model with a cosmological constant has no internal
problems or inconsistencies both at the classical and quantum
levels\cite{Bia10}.

In this regard, the scalar fields such as quintessence
\cite{Cal98,Zla99}, phantom fields \cite{Cal03}, tachyon fields
\cite{Sen02a,Sen02b} and so on, provide a simple dynamical model for
dark energy which can explain cosmic accelerating expansion. The
scalar fields can directly couple to matter Lagrangian, or
indirectly couple to the Ricci scalar
\cite{Dam90,Car92,Car98,Bis06}. If there are negligible
self-interactions for scalar fields, then the experimental bounds on
such fields require it to either couple to the matter much more
weakly than the gravity does, or to couple very strongly
\cite{Che61,Dam02,Uza03,Ber03}. Such fields must be very light
(their mass must be of the order of $H_{0}$, the present Hubble
parameter) in order to evolve cosmologically today. Also, in order
to have consistency with Equivalence Principal, their coupling to
the matter must be extremely small. Chameleon cosmology
\cite{Kho04,Mot04} is a scenario which can address this problem
suitably. In the chameleon cosmology, the scalar fields evolve on
the cosmological time scales today, while according to the
expectations from string theory they have couplings of order unity
to the matter and at the same time remain very light on the
cosmological scales. In fact, the mass of the scalar field is not
constant and instead, it depends on the local matter density. In the
high density regime, the mass of the scalar fields is large, so that
the resulting violations of the Equivalence Principal are
exponentially suppressed. On the cosmological scales, where the
density is very low, the mass of the fields can be of the order of
the present Hubble parameter ($H_{0}$) and cause the current
acceleration of the universe. Such a scalar field is dubbed
``chameleon'' because its physical properties, such as its mass,
depend on its environment \cite{Kho04,Mot04}. We note that this
definition of the ``chameleon'' field is too wide and includes
situations such as plasma density dependence of plasmon mass and
also density and temperature dependence of elementary particles
masses through their effective potentials. In our setup the term
chameleon has sense only in the case of specific exponential
coupling of known quantum fields of matter to the chameleon field
\cite{Dam02}.

The present work has been organized as follows: In section 2, we
introduce the idea of chameleon cosmology. We consider a quintom
scalar field which is non-minimally coupled to the matter (a quintom
field is a hybrid of a quintessence and a phantom field
\cite{Guo05,Zha06b,Laz06,Laz07,Fen05,Fen06}). In this section, by
means of the conservation equation, we derive the matter energy
density which depends on the hybrid scalar field. We study the
effective chameleon potential, its minimum and the mass of two
fields about the minimum of the potential. We show that the value of
the scalar fields and their masses at the minimum depend on local
energy density. In section 3, we study the cosmological dynamics of
the model in the dynamical system approach and we provide a detailed
phase space analysis of the model. We find that in the parameters
space of the model it is possible to have a chameleon dominated
stable attractor which has a negative effective equation of state.
We also show that in this setup, the effective equation of state
parameter of the model crosses the phantom divide line and the
deceleration parameter becomes negative in the past. So this model
has the capability to explain the late-time cosmic speed-up.

\section{The setup}
The action of a two-field chameleon model in 4-dimensions can be
written as follows
\begin{widetext}
\begin{equation}
S=\int
d^{4}x\,\sqrt{-g}\Bigg(\frac{R}{2\kappa^{2}}+\frac{1}{2}g^{\mu\nu}\partial_{\mu}\phi\partial_{\nu}\phi-
\frac{1}{2}g^{\mu\nu}\partial_{\mu}\varphi\partial_{\nu}\varphi+V(\phi,\varphi)+{\cal{L}}_{m}(\psi^{(i)},g_{\mu\nu}^{(i)})\Bigg),
\end{equation}
\end{widetext}
where $\kappa^{2}=\frac{8\pi}{M^2}$ is the gravitational coupling
and ${\cal{L}}_{m}$ represents the Lagrangian density of the matter
fields. Also, the matter fields $\psi^{(i)}$ are coupled to the
scalar fields by the definition
$g_{\mu\nu}^{(i)}=e^{2\kappa\beta_{(i)}(\phi+\varphi)}g_{\mu\nu}$,
where $\beta_{(i)}$ are dimensionless constants. In this paper, we
assume just a single matter energy density component ($\rho_{m}$)
with coupling $\beta$. Also, according to the expectations from
string theory, we allow $\beta$ to be of the order of unity.

In a flat Friedmann-Robertson-Walker background, variation of action
(1) with respect to the metric, leads to the following equations
\begin{equation}
H^{2}=\frac{\kappa^{2}}{6}\dot{\phi}^{2}-\frac{\kappa^{2}}{6}\dot{\varphi}^{2}+\frac{\kappa^{2}}{3}V(\phi,\varphi)
+\frac{\kappa^{2}}{3}\rho_{m}e^{\kappa\beta(\phi+\varphi)},
\end{equation}
\begin{equation}
2\dot{H}+3H^{2}=-\frac{\kappa^{2}}{2}\dot{\phi}^{2}+\frac{\kappa^{2}}{2}\dot{\varphi}^{2}+\kappa^{2}V(\phi,\varphi)
-\kappa^{2}\omega\rho_{m}e^{\kappa\beta(\phi+\varphi)}.
\end{equation}
To derive these equations we have assumed that the matter field of
the universe is a perfect fluid, so that $p_{m}=\omega\rho_{m}$.
Variation of the action (1) with respect to the scalar fields gives
the following equation of motion
\begin{equation}
\ddot{\phi}-\ddot{\varphi}+3H(\dot{\phi}-\dot{\varphi})+\frac{dV}{d\phi}+\frac{dV}{d\varphi}
-2\kappa\beta\omega\rho_{m}e^{\kappa\beta(\phi+\varphi)}=0.
\end{equation}

Equations (2)-(4) give us the energy conservation equation of the
model as follows
\begin{equation}
\dot{\rho}_{m}+3H\rho_{m}(1+\omega)=-\kappa\beta(1+\omega)\rho_{m}(\dot{\phi}+\dot{\varphi}).
\end{equation}
The right hand side of equation (5) shows the non-conservation of
the energy density in this setup, which is due to the presence of
non-minimal coupling between the hybrid scalar field and matter
Lagrangian (in other words, it is due to the presence of the
chameleon field). In the absence of chameleon field (the case that
 $\beta$ goes to zero), equation (5) simplifies to the
ordinary conservation equation.

If we integrate equation (5), we reach the following expression for
the matter energy density
\begin{equation}
\rho_{m}=C\, a^{-3(1+\omega)}\,e^{\kappa\beta(\phi+\varphi)},
\end{equation}
where $C$ is a constant. This equation shows that the scalar fields
and the matter energy density are related to each other via the
chameleon coupling term.

By using equations (2) and (3), we can deduce the effective equation
of state parameter in our setup as follows
\begin{equation}
\omega_{eff}=\frac{p_{eff}}{\rho_{eff}}=\frac{\dot{\phi}^{2}-\dot{\varphi}^{2}-2V(\phi,\varphi)+2\omega
\rho_{m}
e^{\kappa\beta(\phi+\varphi)}}{\dot{\phi}^{2}-\dot{\varphi}^{2}+2V(\phi,\varphi)+2
\rho_{m} e^{\kappa\beta(\phi+\varphi)}}.
\end{equation}

\begin{figure}
\flushleft\leftskip+3em{\includegraphics[width=2.5in]{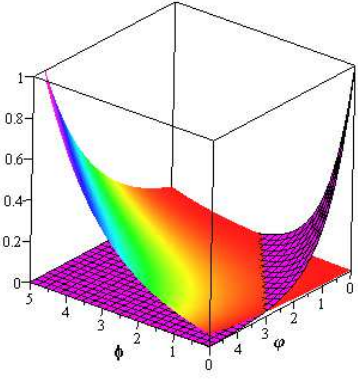}}\hspace{3cm}
\caption{\label{fig:1}The behavior of the runaway potential
(chromatic surface) and the coupling term
$e^{\kappa\beta(\phi+\varphi)}$ (the magenta meshed surface) with
respect to the two components ($\phi$ and $\varphi$) of the hybrid
scalar field. By increment of the scalar fields, the runaway
potential decreases, while the coupling term increases.}
\end{figure}

Equation (4) shows that dynamics of hybrid scalar field in this case
does not depend just only to the $V(\phi,\varphi)$, but instead it
depends on an effective potential which is defined as follows
\begin{equation}
V_{eff}(\phi,\varphi)=V(\phi,\varphi)+\kappa\beta\rho_{m}\,e^{\kappa\beta(\phi+\varphi)},
\end{equation}
which is usually dubbed chameleon effective potential. This
effective potential depends explicitly on the matter density
$\rho_{m}$. We note that we assume a runaway potential defined as
\begin{equation}
V(\phi,\varphi)=V_{0}e^{-\sqrt{6}\kappa(m\phi+n\varphi)}\,,
\end{equation}
which decreases by increment of the scalar fields. Also, the
coupling term ($e^{\kappa\beta(\phi+\varphi)}$) increases as the
scalar fields increase (see Figure ~\ref{fig:1}).

\begin{figure}
\flushleft\leftskip+3em{\includegraphics[width=2.5in]{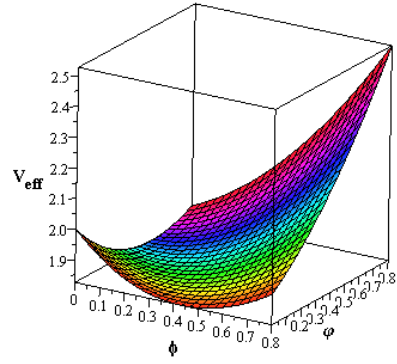}}\hspace{3cm}
\caption{\label{fig:2}The behavior of the effective chameleon
potential with respect to $\phi$ and $\varphi$. The effective
chameleon potential reaches a minimum (a line in this case) during
its evolution.}
\end{figure}

So, if $\beta>0$, $V_{eff}$ has minimum. The behavior of the
effective potential with respect to $\phi$ and $\varphi$ is shown in
figure ~\ref{fig:2}.

We denote the values of $\phi$ and $\varphi$, where the effective
potential becomes minimum (the derivative of the effective potential
becomes zero), by $\phi_{min}$ and $\varphi_{min}$ respectively.
Then, the minimum of the effective potential in our model occurs at
\begin{widetext}
\begin{equation}
(\phi_{min},\varphi_{min})=\Bigg(\phi_{*}\,
,\,\frac{-\ln\left({\frac
{{\beta}^{2}{\kappa}^{2}C\left(-1-\omega+m\right)}{\sqrt{6}m\,a^{3(1+\omega)}}}
\right) +\beta\kappa\phi_{*}\Big(1+\omega\Big)-\beta\kappa
n\phi_{*}-\sqrt{6}n\phi_{*}}{\beta\,\kappa\,
m-\beta\,\kappa\,\omega-\beta\,\kappa+\sqrt{6}m}\Bigg).
\end{equation}
\end{widetext}
The above expression shows that, the minimum of the effective
chameleon potential is a line. Since we deal with positive scalar
field, $\varphi_{min}$ leads us to a constraint on $\phi$ as follows
\begin{equation}
\phi_{*}\geq \frac{\ln
\left(\frac{\beta^{4}\kappa^{4}C^{2}(-1-\omega-m)^{2}}{6m^{2}a^{6(1+\omega)}}\right)}
{2\beta\kappa(1+\omega)-2\beta\kappa\,n-2\sqrt{6}\,n}
\end{equation}
This means that if $\phi$ satisfies the constraint equation (11),
there is a minimum for effective potential as is shown in figure
~\ref{fig:2}.

The mass of the scalar field $\phi$ about the minimum is obtained
from the following relation
\begin{equation}
m_{\phi}^{2}=\frac{\partial^{2}V_{eff}}{\partial\phi^{2}}\Big{|}_{\phi=\phi_{min}}.
\end{equation}
So, we find
\begin{eqnarray}
(m_{\phi})_{min}^{2}=6m^{2}V_{min}+\kappa^{2}\beta^{2}\rho_{m}e^{\kappa\beta(\phi_{min}+\varphi_{min})}\hspace{1cm}\nonumber\\
+\kappa^{2}\beta^{2}(1+\omega^{2})\rho_{m}e^{\kappa\beta(\phi_{min}+\varphi_{min})}.\hspace{1.1cm}
\end{eqnarray}
Also the mass of the scalar field $\varphi$ is given by
\begin{equation}
m_{\varphi}^{2}=\frac{\partial^{2}V_{eff}}{\partial\varphi^{2}}\Big{|}_{\varphi=\varphi_{min}}\,,
\end{equation}
leading to the result
\begin{eqnarray}
(m_{\varphi})_{min}^{2}=6n^{2}V_{min}+\kappa^{2}\beta^{2}\rho_{m}e^{\kappa\beta(\phi_{min}+\varphi_{min})}\hspace{1cm}\nonumber\\
+\kappa^{2}\beta^{2}(1+\omega^{2})\rho_{m}e^{\kappa\beta(\phi_{min}+\varphi_{min})}.\hspace{1.3cm}
\end{eqnarray}
As equations (13) and (15) show, $(m_{\phi})_{min}$ and
$(m_{\varphi})_{min}$ are increasing functions of local matter
density, $\rho_{m}$. This means that, the larger values of matter
density lead to larger values of the chameleon field's mass.

\section{Cosmological dynamics}
In this section, we are going to study cosmological dynamics of the
hybrid chameleon model introduced in previous section in the
framework of dynamical system analysis and phase space trajectories
of the model. In this regard, we should firstly introduce some new
convenient dimensionless variables. These dimensionless quantities
help us to translate our equations of the cosmological dynamics in
the language of the autonomous dynamical system. In our setup, the
dimensionless parameters are defined as follows
\begin{eqnarray}
x=\frac{\kappa\,\dot{\phi}}{\sqrt{6}\,H}\,,\quad
y=\frac{\kappa\,\dot{\varphi}}{\sqrt{6}\,H}\,,\quad
z=\frac{\kappa\,\sqrt{V}}{\sqrt{3}\,H}\,,\quad\hspace{1cm}\nonumber\\
u=\frac{\kappa\,\sqrt{\rho_{m}e^{\kappa\beta(\phi+\varphi)}}}{\sqrt{3}\,H}.\hspace{1cm}
\end{eqnarray}
By rewriting the Friedmann equation (2) in terms of the new
variables, we reach a constraint on the parameters space of the
model as follows
\begin{equation}
1=x^{2}-y^{2}+z^{2}+u^{2},
\end{equation}
by which we can express one of the dimensionless variables in terms
of the others.

By introducing a new time variable $\tau\equiv \ln a$, we obtain the
following autonomous system in our setup
\begin{equation}
\frac{dx}{d\tau}=-3x+3mz^{2}+\frac{\sqrt{6}}{2}\beta\omega
u^{2}+\frac{3x}{2}\Bigg[1+x^{2}-y^{2}-z^{2}+\omega u^{2}\Bigg],
\end{equation}
\begin{equation}
\frac{dy}{d\tau}=-3y-3nz^{2}-\frac{\sqrt{6}}{2}\beta\omega
u^{2}+\frac{3x}{2}\Bigg[1+x^{2}-y^{2}-z^{2}+\omega u^{2}\Bigg],
\end{equation}
\begin{equation}
\frac{dz}{d\tau}=-3mzx-3nzy+\frac{3z}{2}\Bigg[1+x^{2}-y^{2}-z^{2}+\omega
u^{2}\Bigg],
\end{equation}
\begin{eqnarray}
\frac{du}{d\tau}=-\frac{3}{2}(1+\omega)u-\frac{\beta}{2}\omega(\sqrt{6}\,x+\sqrt{6}\,y)u\hspace{1.4cm}\nonumber\\
+\frac{3u}{2}\Bigg[1+x^{2}-y^{2}-z^{2}+\omega u^{2}\Bigg].
\end{eqnarray}

Now, in order to analyze the cosmological evolution in the dynamical
system approach, we should find fixed (or critical) points of the
model. Fixed points are defined as those points that autonomous
equations (18)-(21) all vanish. Eliminating $z$ by using constraint
equation (17), we obtain the critical points of our setup. We find
four critical points (\textbf{M}, \textbf{N}, \textbf{P},
\textbf{Q}) and two critical lines (${\cal{L}}_{1}$ and
${\cal{L}}_{2}$) in our setup which we have summarized their
properties in tables ~\ref{tab:table1}, ~\ref{tab:table2} and
~\ref{tab:table3}.

\begin{table*}
\caption{\label{tab:table1}Location and existence of critical points
and corresponding effective equation of state parameter. $m_{*}$
and ($x_{Q}$,$y_{Q}$,$u_{Q}$) are defined by equations (22) and (23)
respectively.}
\begin{ruledtabular}
\begin{tabular}{ccccc}
Point & ($x$,$y$,$u$) & existence & $\omega_{eff}$\\ \hline
\textbf{M} & ($m$,$-n$,$0$) &
all $m$ and $n$ and all $\omega$ & $2m^{2}-2n^{2}-1$ \\
\textbf{N} &
($\frac{\sqrt{6}\beta\omega}{3-3\omega},\frac{\sqrt{6}\beta\omega}{3\omega-3},1$)
 & all $m$ and $n$ and $\omega\neq 1$ & $\omega$ \\
\textbf{P} &
($\frac{\sqrt{6}\beta\omega}{3-3\omega},\frac{\sqrt{6}\beta\omega}{3\omega-3},-1$)
 & all $m$ and $n$ and $\omega\neq 1$ & $\omega$ \\
\textbf{Q} & ($x_{Q}$,$y_{Q}$,$u_{Q}$) & $m\geq m_{*}$ and all
$\omega$ & $\frac{x_{Q}^{2}-y_{Q}^{2}-z_{Q}^{2}+\omega
u_{Q}^{2}}{x_{Q}^{2}-y_{Q}^{2}+z_{Q}^{2}+u_{Q}^{2}}
$ \\
\end{tabular}
\end{ruledtabular}
\end{table*}

Now we discuss characters of each critical point separately.
\begin{itemize}
{\item{\textbf{Critical Point M}:}

Point \textbf{M} represents either a solution with a scalar field's
kinetic energy term domination or potential energy term domination,
depending on the values of $m$ and $n$. Also, depending on these
values, this solution can be stable or unstable. For instance, by
taking $m=0.6$ and $n=0.5$, the universe with this solution is an
attractor, potential energy term dominated, meaning that if the
universe reaches this state, it remains there forever. The value of
effective equation of state parameter corresponding to this solution
is negative (-0.78), so this case is corresponding to an
accelerating universe. Figure ~\ref{fig:3} shows the phase space
trajectories of the model in two dimensions for $\beta=1$, $m=0.6$,
$n=0.5$ and $\omega=0$ (corresponding to potential energy
domination). The point \textbf{M} is shown as an attractor point in
this plot. Also, we can see this stable point in 3-dimensional plot
as shown in figure ~\ref{fig:4}. If we set $n=0.7$ and $m=1.2$, the
solution is a saddle, kinetic energy domination. The universe during
its evolution can reach this state but doesn't remain there and
evolves to another state. In this case, the corresponding value of
the effective equation of state parameter is positive
($\omega_{eff}=0.9$) and shows a decelerating universe. This saddle
point of the phase space trajectories is shown in figure
~\ref{fig:5}. This figure is plotted for $n=0.7$ and $m=1.2$. The
point \textbf{M} would be corresponding to a repeller, kinetic term
dominated solution if for instance we set $m=1.6$, $n=0.8$ and
$\omega=\frac{1}{3}$ (see figure ~\ref{fig:6}). This unstable
solution is relevant to early times cosmology.}

\begin{figure}
\flushleft\leftskip+3em{\includegraphics[width=2.5in]{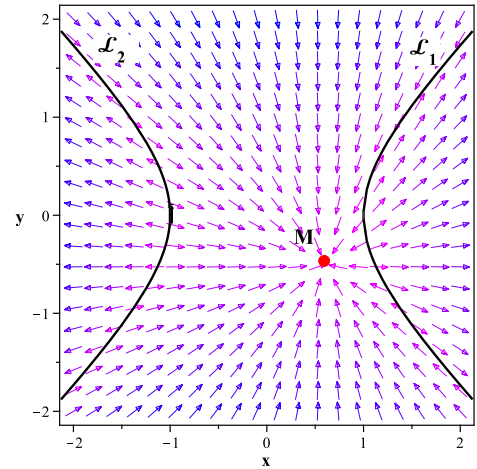}}\hspace{3cm}
\caption{\label{fig:3}The phase space trajectories of the hybrid
chameleon model with $m=0.6$, $n=0.5$, $\beta=1$ and $\omega=0$.
This figure is plotted for the case with $u=0$. With these values in
the parameters space, the critical lines ${\cal{L}}_{1}$ and
${\cal{L}}_{2}$ (the solutions corresponding to the quintessence
component's kinetic energy domination) are repeller (and so,
unstable solutions). The critical point \textbf{M} (the solution
with potential energy domination) is a stable attractor.}
\end{figure}

\begin{table*}
\caption{\label{tab:table2}Eigenvalues and dynamical characters of
the fixed points.}
\begin{ruledtabular}
\begin{tabular}{ccccc}
Point & Eigenvalues($\lambda_{1}$,\,$\lambda_{2}$,\,$\lambda_{3}$) & stability \\
\hline \textbf{M} &
\Big($-\frac{3}{2}-\frac{3}{2}\omega-\frac{\sqrt{6}\beta\omega
m-\sqrt{6}\beta\omega n}{2}+3m^{2}-3n^{2}$, & stable if
$m<\frac{(-2n\beta^{2}\sqrt{n^{2}+1}+2\sqrt{6}\beta\sqrt{n^{2}+1})
\sqrt {6}}{( 3n+\sqrt{6}\beta-3\sqrt{n^{2}+1})(-2n\beta+\sqrt {6}-2\beta\sqrt{n^{2}+1})}$ \\
&$-3+3m^{2}-3n^{2}$,\,$-3+3m^{2}-3n^{2}$\Big)& $+\frac{ (
-3\,{n}^{2}-3-2\,{n}^{2}{\beta}^{2}-2\,{\beta}^{2}+3\, n\sqrt
{{n}^{2}+1})\sqrt{6}}{( 3\,n+\sqrt {6}\beta-3\, \sqrt
{{n}^{2}+1})( -2\,n\beta+\sqrt {6}-2\,\beta\, \sqrt {{n}^{2}+1}
)}$\\\\
\textbf{N} & \Big(${\frac
{-3+3\omega^{2}+2m\sqrt{6}\beta\omega-2n\sqrt{6}\beta\omega}{-1+\omega}}
$, & stable if $\omega<1$ and
$m<\frac{(3-3\omega^{2}+2n\sqrt{6}\beta\omega)\sqrt{6}}{12\beta\omega}
$ \\
&$-\frac{3}{2}+\frac{3}{2}\omega$,\,$-\frac{3}{2}+\frac{3}{2}\omega$\Big)& saddle point if
$\omega=0$\\\\
\textbf{P} & \Big(${\frac
{-3+3\omega^{2}+2m\sqrt{6}\beta\omega-2n\sqrt{6}\beta\omega}{-1+\omega}}
$,
 &  stable if $\omega<1$ and $m<\frac{(3-3\omega^{2}+2n\sqrt{6}\beta\omega)\sqrt{6}}{12\beta\omega}
$  \\
&$-\frac{3}{2}+\frac{3}{2}\omega$,\,$-\frac{3}{2}+\frac{3}{2}\omega$\Big)& saddle point if
$\omega=0$\\\\
\textbf{Q} & ($\lambda_{1Q}$,\,$\lambda_{2Q}$,\,$\lambda_{3Q}$) & stable \\
\end{tabular}
\end{ruledtabular}
\end{table*}

\begin{figure}
\flushleft\leftskip+3em{\includegraphics[width=2.5in]{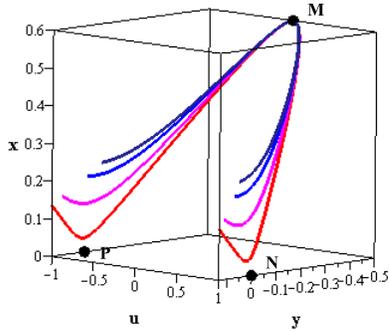}}\hspace{3cm}
\caption{\label{fig:4}3-dimensional phase space trajectories of the
model with $m=0.6$, $n=0.5$, $\beta=1$ and $\omega=0$. For these
values of parameters, we have a potential energy dominated solution
(point {\textbf{M}}) which is a stable attractor. The points
{\textbf{N}} and {\textbf{P}} (corresponding to the effectively
matter dominated era) are saddle points.}
\end{figure}

\begin{figure}
\flushleft\leftskip+3em{\includegraphics[width=2.5in]{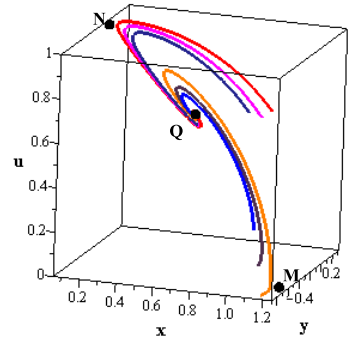}}\hspace{3cm}
\caption{\label{fig:5}3-dimensional phase space trajectories of the
model with $m=1.2$, $n=0.7$, $\beta=1$ and $\omega=0$. For these
values of parameters, the potential energy dominated solution (point
{\textbf{M}}) is a saddle point, while the point {\textbf{Q}}
(corresponding to the chameleon dominated solution) is a stable
attractor solution.}
\end{figure}

\begin{figure}
\flushleft\leftskip+3em{\includegraphics[width=2.5in]{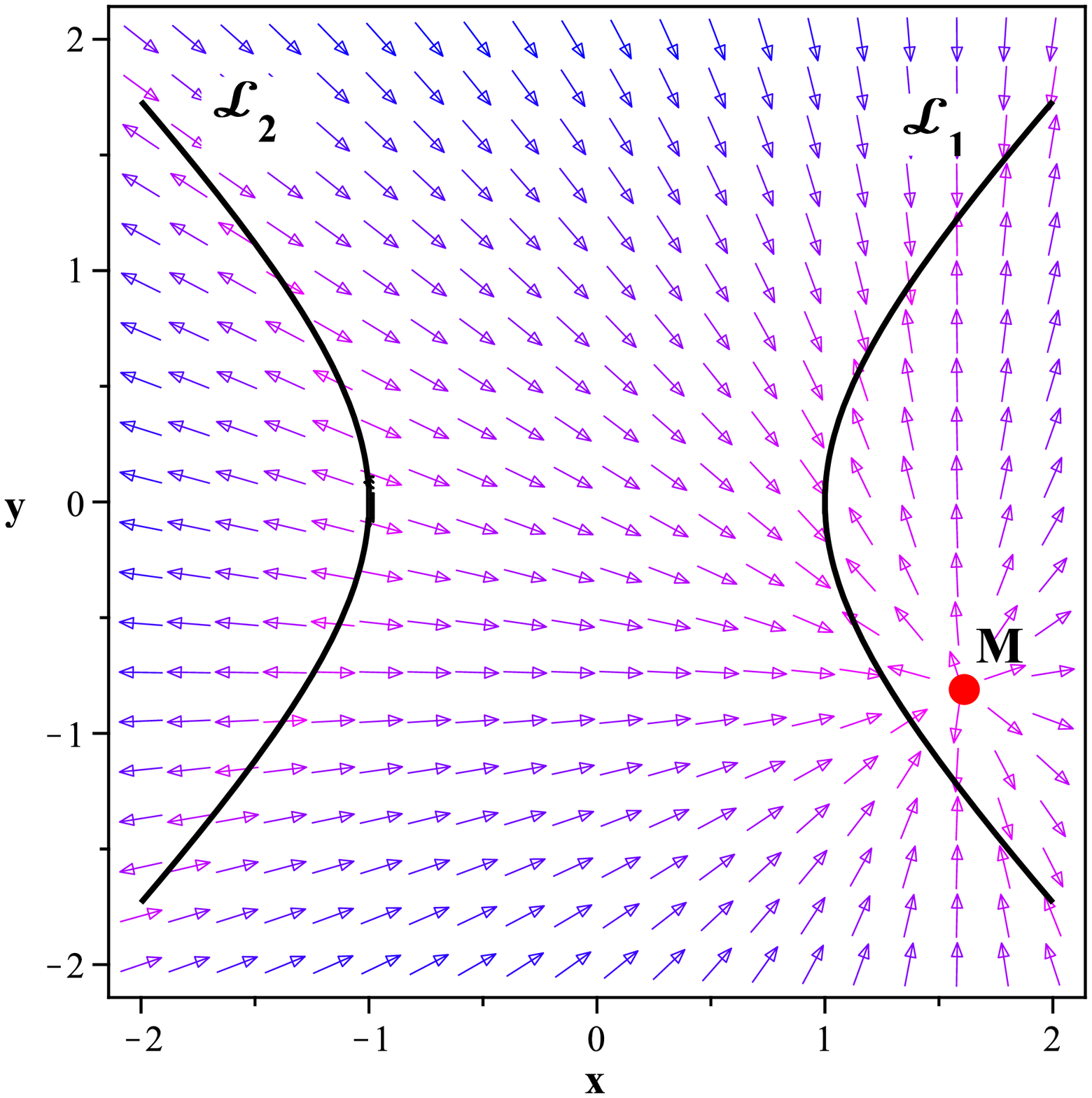}}\hspace{3cm}
\caption{\label{fig:6}The phase space trajectories of the model with
$m=1.6$, $n=0.8$, $\beta=1$ and $\omega=\frac{1}{3}$. Like as figure
3, this figure is plotted also for the case with $u=0$. With these
values of parameters, in contrast to figure 3, the critical line
${\cal{L}}_{1}$ is stable attractor, while the critical line
${\cal{L}}_{2}$ remains unstable. Also, in this case the critical
point \textbf{M} is a repeller.}
\end{figure}

{\item{\textbf{Critical Point N}:}

Critical point \textbf{N} exists if $\omega\neq 1$ (in other words,
if there is no stiff fluid in the universe). For this solution, the
effective equation of state parameter is exactly the same as the
equation of state parameter of the matter and so the universe with
this solution is effectively matter dominated. For $\omega<1$,
depending on the values of $m$ and $n$, this matter dominated
solution can be stable or saddle point. Also, with $\omega=0$, this
solution is always a saddle, independent of the values of $m$ and
$n$. This means that the universe, during its evolution, experiences
this matter domination state and then evolves to another state. The
point \textbf{N}, in both figures ~\ref{fig:4} and ~\ref{fig:5}, is
a saddle point in the phase space of the model. It should be noticed
that since both figures have been plotted with $\omega=0$, for both
$m=0.6$ and $m=1.1$, \textbf{N} is a saddle point.}

{\item{\textbf{Critical Point P}:}

Like as the critical point \textbf{N}, the critical point \textbf{P}
exists if $\omega\neq 1$. The properties of this critical point is
the same as the properties of the critical point \textbf{N}. In
figure ~\ref{fig:4}, we can see the point \textbf{P} as a saddle
point in the phase space trajectories of the model. This point is
not shown in figure ~\ref{fig:3}, because this figure has been
plotted for positive $u$ while \textbf{P} is located at $u=-1$.}

{\item{\textbf{Critical Point Q}:}

The critical point \textbf{Q} exists if there is the following
constraint on the parameters space of the model
\begin{widetext}
\begin{eqnarray}
m>\frac{\sqrt{6}}{12}\beta\omega+
\frac{\sqrt{6\beta^{2}\omega^{2}+72+144n^{2}+72\omega-24\sqrt{6}\beta\omega
n}}{12} \equiv m_{*}\,.
\end{eqnarray}
\end{widetext}

If this critical point exists, its location is at
\begin{equation}
(x_{Q},\,y_{Q},\,u_{Q})
\end{equation}
where
\begin{widetext}
$$
x_{Q}=\frac{12n^{2}\beta\omega-3m\sqrt{6}+6\omega\beta-12mn\beta\omega+3m\sqrt{6}\omega^{2}+6\omega^{2}\beta
-2\sqrt{6}\beta^{2}\omega^{2}n+2\sqrt{6}\beta^{2}\omega^{2}m}{\sqrt{6}(-6m\omega+6m-6n\omega+6n+\sqrt{6}\omega^{2}\beta
-3\sqrt{6}\omega\beta)(n-m)}\,,
$$
$$
y_{Q}=\frac{6\beta\omega+6\beta\omega^{2}-2\sqrt{6}\beta^{2}\omega^{2}n-3\sqrt{6}n
+3\sqrt{6}n\omega^{2}+12m\beta\omega n+2\sqrt{6}\beta^{2}\omega^{2}m
-12m^{2}\omega\beta}{\sqrt{6}(6n+6m-6n\omega-6m\omega-3\beta\omega\sqrt{6}+\beta\omega^{2}\sqrt{6})
(m-n)}\,,
$$
and
$$
u_{Q}=\frac{\sqrt{3n+3m-2\beta\omega\sqrt{6}-6n\omega-6m\omega+2\sqrt{6}\omega^{2}\beta+3n\omega^{2}
+3\omega^{2}m-2\beta^{2}\omega^{2}n+2\beta^{2}\omega^{2}m}}{\left(
-6\,m\omega+6 \,m-6\,n\omega+6\,n+\sqrt
{6}{\omega}^{2}\beta-3\,\beta\,\omega\, \sqrt {6}
\right)(n-m)^\frac{1}{2}}
$$
$$\times\sqrt{-12m^{2}+6+12n^{2}+6\omega+2\sqrt{6}m\omega\beta-2\sqrt{6}\omega\beta
n}.$$ \end{widetext}

Also, $(\lambda_{1Q}, \lambda_{2Q}, \lambda_{3Q})$ are the
eigenvalues of the following Matrix

\begin{equation}
M=\left(\begin{array}{lccr} \frac{\partial x'}{\partial x} &
\frac{\partial x'}{\partial y} & \frac{\partial x'}{\partial u}\\
\frac{\partial y'}{\partial x} & \frac{\partial y'}{\partial y} &
\frac{\partial y'}{\partial u} \\
\frac{\partial u'}{\partial x} & \frac{\partial u'}{\partial y} &
\frac{\partial u'}{\partial u}
\end{array}
\right)_{(x,y,u)=(x_{Q},y_{Q},u_{Q})},
\end{equation}
where prime refers to derivative with respect to $\tau$ (see
equations (18)-(21)). We note that since the eigenvalues of this
solution are so lengthy and complicate, we avoid to express them
here explicitly. The universe with this solution experiences an
accelerating phase if

\begin{eqnarray}
-\sqrt{y_{Q}^{2}+\frac{z_{Q}^{2}}{2}-u_{Q}^{2}(\omega+\frac{1}{3})}\leq
x_{Q} \hspace{2cm}\nonumber\\
\leq
\sqrt{y_{Q}^{2}+\frac{z_{Q}^{2}}{2}-u_{Q}^{2}(\omega+\frac{1}{3})}.
\end{eqnarray}

Point \textbf{Q} of figure ~\ref{fig:5} is a stable chameleon
dominated solution and this means that if the universe reaches this
state, remains there forever. In this solution, the universe can
experience the late time acceleration. Figure ~\ref{fig:5} shows
3-dimensional phase space of our setup with $\beta=1$, $\omega=0$,
$m=1.2$ and $n=0.7$. With these values of the parameters, the
critical points \textbf{M} and \textbf{N} are saddle points and the
points \textbf{Q} is a stable attractor. In summary, we can say that
with these parameters values, the universe during its evolution
reaches a kinetic energy dominated era and then evolves to a matter
dominated era. After that the universe evolves to a chameleon
dominated era and remains there forever. Also, with this choice of
values, the value of the effective equation of state parameter is
about $-1.1$. This means that the universe in this parameters space
experiences late time acceleration and its stable state lies in
phantom-like phase.}

{\item{{\bf Critical Line}\,\,\large{${\cal_{L}}_{1}$}:}

In this model, we also have two critical lines. The critical line
${\cal{L}}_{1}$, which is located at $(x_{*},\,
\sqrt{x_{*}^{2}-1},\,\,0)$, is a quintessence's kinetic term
dominated solution. So, there is a constraint on this solution. This
kinetic energy dominated solution exists if

\begin{equation}
x_{*}\geq 1 \quad\quad or \quad\quad x_{*}\leq 1\,.
\end{equation}

The stability of this line depends on the model parameters values.
In the parameters space which satisfies the following constraint,
this solution is a stable attractor
\begin{widetext}
$$
m>\,-\frac {\sqrt {6}n\beta\,\omega\,{x}^{2}+\sqrt {6}x\sqrt
{-1+{x}^{2}} n\beta\,\omega-\beta\,\omega\,\sqrt {6}x-\sqrt
{6}n\beta\,\omega-\beta \,\omega\,\sqrt {6}\sqrt {-1+{x}^{2}}}{x
\left( \beta\,\omega\,\sqrt {6}x+ \beta\,\omega\,\sqrt {6}\sqrt
{-1+{x}^{2}}+3\,\omega-3 \right)}
$$
\begin{equation}
+\frac{+3+3\,n\sqrt {-1+{x}^{2}}\omega-3 \,n\sqrt
{-1+{x}^{2}}-3\,\omega}{x \left( \beta\,\omega\,\sqrt {6}x+
\beta\,\omega\,\sqrt {6}\sqrt {-1+{x}^{2}}+3\,\omega-3
\right)}\equiv m_{{\cal{L}}_{1}}\,,
\end{equation}
\end{widetext}

In figure ~\ref{fig:3} where the chosen values of the parameters do
not satisfy the constraint equation (27), the critical line
${\cal{L}}_{1}$ is unstable. But in figure ~\ref{fig:6} this line is
a stable attractor because the values of parameters used to plot
this figure satisfy the mentioned constraint. }

\begin{table*}
\caption{\label{tab:table3}Location, eigenvalues and dynamical
characters of the critical lines. Note that, $m_{{\cal{L}}_{1}}$ is
defined in (27).}
\begin{ruledtabular}
\begin{tabular}{cccccc}
Line & ($x$,$y$,$u$) & existence &Eigenvalues& $\omega_{eff}$&stability\\
\hline ${\cal{L}}_{1}$ & ($x_{*}$,$\sqrt{x_{*}^{2}-1}$,$0$) & all
$m$,\, $n$ and $\omega$ if $x\geq 1$
&($\lambda_{1{\cal{L}}_{1}}$,\,$\lambda_{2{\cal{L}}_{1}}$,\,
$\lambda_{3{\cal{L}}_{1}}$)& $1$ &stable if \\
&&&&&$m>m_{{\cal{L}}_{1}}$\\\\
${\cal{L}}_{2}$ & ($x_{*}$,$-\sqrt{x_{*}^{2}-1}$,$0$)
 & all $m$,\, $n$ and $\omega$ if $x\leq -1$ &($\lambda_{1{\cal{L}}_{2}}$,\,$\lambda_{2{\cal{L}}_{2}}$,\,
$\lambda_{3{\cal{L}}_{2}}$)& $1$& unstable\\
\end{tabular}
\end{ruledtabular}
\end{table*}

{\item{{\bf Critical Line}\,\,\large{${\cal_{L}}_{2}$}:}

This phantom's kinetic energy dominated critical line is located at
$(-x_{*},\,\, -\sqrt{x_{*}^{2}-1},\,\, 0)$. So, for existence of
this solution the constraint equation (27) should be satisfied. In
contrast to ${\cal{L}}_{1}$ case, the critical line ${\cal{L}}_{2}$
is an unstable solution for all values of the model parameters space
(see figures ~\ref{fig:3} and ~\ref{fig:6}). }

\end{itemize}

\begin{figure}
\flushleft\leftskip+3em{\includegraphics[width=2.5in]{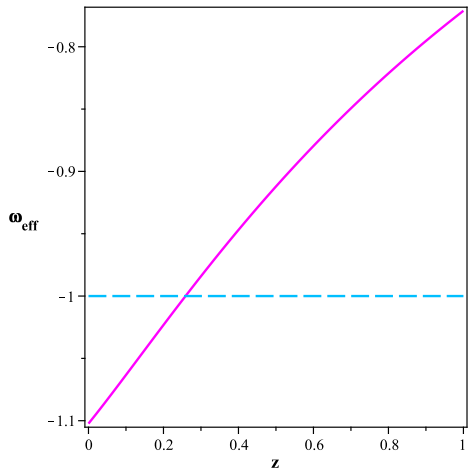}}\hspace{3cm}
\caption{\label{fig:7}The evolution of the effective equation of
state parameter of the hybrid chameleon model versus the red-shift
for the case that $m=1.2$, $n=0.7$, $\beta=1$ and $\omega=0$.}
\end{figure}

A parameter which gives us a suitable background to understand the
dynamics of the universe, the nature of dark energy and the
possibility of crossing of the phantom-divide line, is the equation
of state parameter. According to the recent observational data, the
equation of state parameter of the dark energy crossed the phantom
divide line ($\omega=-1$) in the near past. It is shown that,
considering a dynamical dark energy component in a cosmological
setup enables the model to explain these observational evidence
\cite{Cop06,Noj04,Cal02,Ark02,Pia04,Wei06,Vik05,Ani05,Wan05,Noj05,Noj06,Eli05,Zha06a,Apo06,Ala04,Nes05,lib07,Sch08,Bri07,Sam09,Cal09,Sah04,Sah06}.
Figure ~\ref{fig:7} shows the evolution of $\omega_{eff}$ with
respect to the red-shift parameter in our setup. This figure has
been plotted by adopting the ansatz with $a=a_{0}\, e^{\beta t}$,
$\varphi=\varphi_{0}\, e^{-\delta t}$ and $\phi=\phi_{0}\,
e^{-\alpha t}$ ($\alpha$, $\delta$ and $\beta$ are positive
constants). Note that in this figure, the values of parameters are
the same as figure ~\ref{fig:5} (where the chameleon dominated
solution is a stable attractor). Also, we have set $a_{0}=1$,
$\phi_{0}=0.8$, $\alpha=1$, $\delta=1.6$ and $\beta=1$. As figure
shows, in the presence of the chameleon field, the universe enters
the phantom phase in the near past at $z\simeq 0.26$. So, in this
model, the universe experiences a smooth crossing of the phantom
divide, $\omega_{eff}=-1$, line.

\begin{figure}
\flushleft\leftskip+3em{\includegraphics[width=2.5in]{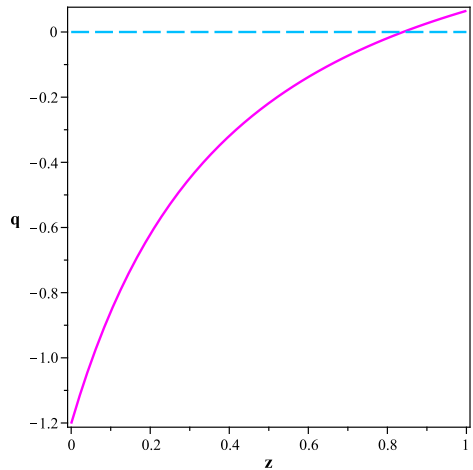}}\hspace{3cm}
\caption{\label{fig:8}The evolution of the deceleration parameter of
the hybrid chameleon model versus the red-shift for the case that
$m=1.2$, $n=0.7$, $\beta=1$ and $\omega=0$.}
\end{figure}

The deceleration parameter, $q$, is another important parameter in
cosmological evolution. Its positive value (corresponding to
$\ddot{a}<0$), shows the decelerating expansion of the universe and
its negative value (corresponding to $\ddot{a}>0$), shows the
accelerating expansion. $q$ is defined as follows
\begin{equation}
q=-\frac{\dot{H}}{H^{2}}-1.
\end{equation}
The evolution of the deceleration parameter versus red-shift is
plotted in figure ~\ref{fig:8}. We can see from this figure that the
deceleration parameter has become negative in the past at $z\simeq
0.84$, meaning that the universe has entered to an accelerating
phase at $z\simeq 0.84$. So, a hybrid chameleon model has the
phantom like behavior and can explain the late time cosmic
acceleration of the universe in an observationally viable manner.

\section{Summary}

In the current paper, we have studied a hybrid chameleon model in
details. By choosing a runaway potential and by using of the
coupling term $e^{\kappa\beta(\phi+\varphi)}$, we have introduced an
effective chameleon potential. In contrast with runaway potential
and coupling term, this effective chameleon potential has a minimum.
We have derived this minimum and the messes of the hybrid chameleon
field in the minimum. We have found that the minimum of the
effective chameleon potential is a line and in order to existence
this minimum, there is a constraint on the parameters space of the
model. It has been shown that the chameleon field and the matter
energy density is related to each other. Also, we have shown that
the larger values of the density leads to larger values of the
chameleon field's masses. Then we have considered the cosmological
dynamics of the hybrid chameleon model. In a dynamical system
approach, we have studied the phase space trajectories of the model
and its stability. We have found four critical points and two
critical lines in our setup and have derived their eigenvalues in
order to find the stable solutions. The critical lines and points
can be attractor, repeller and saddle, depending on the values of
the parameter spaces. By analyzing two and three dimensional phase
spaces, it was shown that there are matter domination, scalar
field's kinetic term domination and chameleon domination stable
solutions, depending on the values of the parameters space. With
some values of the parameters space, we have found the solution
where the universe after passing the scalar field's kinetic term
dominated era (repeller) and the potential energy dominated and
matter dominated era (saddle), reach a stable attractor chameleon
dominated era and remains there forever (one can see the saddle
point and attractor in figure ~\ref{fig:5}). Also, the effective
equation of state parameter in this case is below $-1$. By adopting
the parameter's values used in figure ~\ref{fig:5}, we have analyzed
the late time dynamics of the hybrid chameleon model. We have shown
that the universe in this setup has crossed the phantom divide line
at $z\simeq 0.26$ and has entered to the accelerating phase at
$z\simeq 0.84$. So, a hybrid chameleon model can explain the late
time cosmic acceleration and realize the phantom behavior.


\begin{thebibliography}{}

\bibitem[Riess (1998)]{Rie98} A. G. Riess \emph{et al}., Astron. J. \textbf{116} 1009 (1998).

\bibitem[Perlmutter (1999)]{Per99} S. J. Perlmutter \emph{et al}., Astrophys. J. \textbf{517} 565 (1999).

\bibitem[Riess (2004)]{Rie04} A. G. Riess {\it et al.}, Astrophys. J. {\bf607} 665  (2004).

\bibitem[Astier (2006)]{Ast06} P. Astier {\it et al.}, Astron. Astrophys. {\bf447} 31 (2006).

\bibitem[Wood-Vasey (2007)]{Woo07} W. M. Wood-Vasey {\it et al.}, Astrophys. J. {\bf666}
694-715 (2007).

\bibitem[Spergel (2007)]{Spe07} D. N. Spergel {\it et al.}, Astrophys. J. Suppl {\bf 170} 377 (2007).

\bibitem[Hinshaw (2007)]{Hin07} G. Hinshaw {\it et al.}, Astrophys. J. Suppl , {\bf288} 170 (2007).

\bibitem[Colless (2001)]{Col01} M. Colless {\it et al}, Mon. Not. R. Astron. Soc. {\bf328} 1039 (2001).

\bibitem[Tegmark (2004)]{Teg04} M. Tegmark {\it et al.}, Phys. Rev. D {\bf69} 103501 (2004).

\bibitem[Cole (2005)]{Col05} S. Cole {\it et al}., Mon. Not. R. Astron. Soc. {\bf 362} 505 (2005).

\bibitem[Springel (2006)]{Spr06} V. Springel, C. S. Frenk and S. M. D. White, Nature (London)
{\bf440} 1137 (2006).

\bibitem[Boughn (2004)]{Bou04} S. P. Boughn and R. G. Crittenden, {\it Nature} {\bf427} 24 (2004).

\bibitem[McEwen (2007)]{McE07} J. D. McEwen {\it et al.}, Mon. Not. R. Astron. Soc. {\bf376}
1211 (2007).

\bibitem[Komatsu (2009)]{Kom09} E. Komatsu {\it et al.} [WMAP Collaboration], Astrophys. J. Suppl.
{\bf180} 330 (2009).

\bibitem[Copeland (2006)]{Cop06} E. J. Copeland, M. Sami and S. Tsujikawa,  Int. J. Mod. Phys. D
{\bf15} 1753 (2006).

\bibitem[Nojiri (2004)]{Noj04} S. Nojiri and S. D. Odintsov, Phys. Rev. D {\bf70} 103522 (2004).

\bibitem[Caldwell (2002)]{Cal02} R. R. Caldwell, Phys. Lett. B {\bf545} 23 (2002).

\bibitem[Arkani-Hamed (2002)]{Ark02} N. Arkani-Hamed, P. Creminelli, S. Mukohyama and M. Zaldarriaga,
JCAP {\bf0404} 001 (2004).

\bibitem[Piazza (2004)]{Pia04} F. Piazza and S. Tsujikawa, JCAP {\bf0407} 004 (2004).

\bibitem[Wei (2006)]{Wei06} H. Wei and R. G. Cai, \prd {\bf73} 083002 (2006).

\bibitem[Vikman (2005)]{Vik05} A. Vikman, \prd {\bf71} 023515 (2005).

\bibitem[Anisimov (2005)]{Ani05} A. Anisimov, E. Babichev and A. Vikman, JCAP {\bf0506} 006 (2005).

\bibitem[Wang (2005)]{Wan05} B. Wang, Y.G. Gong and E. Abdalla, Phys. Lett. B {\bf 624}
141 (2005).

\bibitem[Nojiri (2005)]{Noj05} S. Nojiri, S. D. Odintsov and S. Tsujikawa, \prd {\bf71}
(2005) 063004.

\bibitem[Nojiri (2006)]{Noj06} S. Nojiri and S. D. Odintsov, Gen. Rel. Grav, {\bf38} 1285 (2006).

\bibitem[Elizalde (2005)]{Eli05} E. Elizalde, S. Nojiri, S. D. Odintsov and P. Wang,
\prd {\bf 71} 103504 (2005).

\bibitem[Zhao (2006a)]{Zha06a} W. Zhao and Y. Zhang, \prd {\bf 73} 123509 (2006a).

\bibitem[Apostolopoulos (2006)]{Apo06} P. S. Apostolopoulos and N. Tetradis, \prd {\bf 74}
064021 (2006).

\bibitem[Alam (2004)]{Ala04} U. Alam , V. Sahni and A. A. Starobinsky, JCAP {\bf06}
008 (2004).

\bibitem[Nesseris (2005)]{Nes05} S. Nesseris and L. Perivolaropoulos, \prd {\bf72}
123519 (2005).

\bibitem[Libanov (2007)]{lib07} M. Libanov, E. Papantonopoulos, V. Rubakov, M. Sami and S.
Tsujikawa, JCAP {\bf0708} 010 (2007).

\bibitem[Scherrer (2008)]{Sch08} R. J. Scherrer and A. A. Sen, \prd {\bf 78} 067303 (2008).

\bibitem[Briscese (2007)]{Bri07} F. Briscese, E. Elizalde, S. Nojiri and S. D. Odintsov, Phys. Lett.
B {\bf646} 105 (2007).

\bibitem[Sami (2009)]{Sam09} M. Sami, [arXiv:0901.0756] (2009).

\bibitem[Caldera-Cabral (2009)]{Cal09} G. Caldera-Cabral, R. Maartens, L. A. Urena-Lopez, Phys. Rev. D
{\bf79} 063518 (2009).

\bibitem[Sahni (2004)]{Sah04} V. Sahni, Lect. Notes Phys. {\bf653} 141 (2004).

\bibitem[Sahni (2006)]{Sah06} V. Sahni and A. Starobinsky, Int. J. Mod. Phys. D {\bf15}
2105 (2006).

\bibitem[Padmanabhan (2003)]{Pad03} T. Padmanabhan, Phys. Rept. {\bf380} 235 (2003).

\bibitem[Carroll (2001)]{Car01} S. M. Carroll, Living Rev. Rel. {\bf4} 1 (2001).

\bibitem[Sahni (2000)]{Sah00} V. Sahni and A. Starobinsky,
Int. J. Mod. Phys. D \textbf{9} 373 (2000).

\bibitem[Peebles (2003)] {Pee03} P. J. E. Peebles and B. Ratra, Rev. Mod. Phys. \textbf{75} 559
(2003).

\bibitem[Bianchi (2010)]{Bia10} E. Bianchi and C. Rovelli,
[arXiv:1002.3966].

\bibitem[Caldwell (1998)]{Cal98} R. R. Caldwell, R. Dave and P. J. Steinhardt, \prl \textbf{80}
1582 (1998).

\bibitem[Zlatev (1999)]{Zla99} I. Zlatev, L. Wang and P. J. Steinhardt, \prl {\bf82} 896 (1999).

\bibitem[Caldwell (2003)]{Cal03} R. R. Caldwell, M. Kamionkowski, N. N. Weinberg, \prl
{\bf91} 071301 (2003).

\bibitem[Sen (2002a)]{Sen02a} A. Sen, JHEP {\bf0207} 065 (2002a).

\bibitem[Sen (2002b)]{Sen02b} A. Sen, Mod. Phys. Lett. A {\bf17} 1797 (2002b).

\bibitem[Damour (1990)]{Dam90} T. Damour, G. W. Gibbons and C. Gundlach, \prl {\bf 64}
123 (1990).

\bibitem[Carroll (1992)]{Car92} S. M. Carroll, W. H. Press and E. L. Turner, Ann. Rev. Astron.
Astrophys. {\bf 30} 499 (1992).

\bibitem[Carroll (1998)]{Car98} S. M. Carroll, \prl {\bf 81} 3067 (1998).

\bibitem[Biswas (2006)]{Bis06} T. Biswas, R. Brandenberger, A. Mazumdar and T. Multamaki. \prd {\bf 74} 063501 (2006).

\bibitem[Chew (1961)]{Che61} G. F. Chew and S. C. Frautschi. \prl {\bf 7} 394 (1961).

\bibitem[Damour (2002)]{Dam02} T. Damour, F. Piazza and G. Veneziano, \prd {\bf 66} 046007 (2002).

\bibitem[Uzan (2003)]{Uza03} J. P. Uzan, Rev. Mod. Phys. {\bf 75} 403 (2003).

\bibitem[Bertotti (2003)]{Ber03} B. Bertotti et al. Nature {\bf 425} 374 (2003).

\bibitem[Khoury (2004)]{Kho04} J. Khoury and A. Weltman, \prl {\bf 93} 171104 (2004).

\bibitem[Mota (2004)]{Mot04} D. F. Mota, J. D. Barrow, Phys. Lett. B {\bf 581} 141 (2004).

\bibitem[Guo (2005)]{Guo05} Z. K. Guo, Y. S. Piao, X. M. Zhang and Y. Z. Zhang, Phys. Lett. B
{\bf 608} 177 (2005).

\bibitem[Zhang (2006b)]{Zha06b} X. F. Zhang, H. Li, Y. S. Piao and X. M. Zhang, Mod. Phys. Lett. A
{\bf 21} 231 (2006b).

\bibitem[Lazkoz (2006)]{Laz06} R. Lazkoz and G. Le\'{o}n, Phys. Lett. B {\bf 638} 303 (2006).

\bibitem[Lazkoz (2007)]{Laz07} R. Lazkoz, G. Le\'{o}n and I. Quiros, Phys. Lett. B {\bf 649} (2007)
103.

\bibitem[Feng (2005)]{Fen05} B. Feng, X. L. Wang and X. M. Zhang, Phys. Lett. B {\bf 607} 35 (2005).

\bibitem[Feng (2006)]{Fen06} B. Feng, M. Li, Y. S. Piao and X. Zhang, Phys. Lett. B {\bf 634} 101 (2006).



\end{thebibliography}
\end{document}